\documentclass[prl,twocolumn,showpacs,preprintnumbers]{revtex4}
\usepackage{amssymb}

\usepackage{graphicx}
\usepackage{dcolumn}
\usepackage{bm}

\newcommand{\be}{\begin{equation}}
\newcommand{\ee}{\end{equation}}

\newcommand{\e}{\mbox{$\epsilon$}}

\newcommand{\n}{\mbox{$N_i$}}
\newcommand{\dd}{\mbox{$D_i(\{N_j\})$}}

\newcommand{\nee}{\mbox{$\bar{n}(\e_i)$}}
\newcommand{\nega}{\mbox{$\bar{n}_{\gamma}(\e_i)$}}
\newcommand{\nep}{\mbox{$\bar{n}_p(\e_i)$}}
\newcommand{\negg}{\mbox{$\bar{n}_g(\e_i)$}}

\newcommand{\bfg}{\begin{figure}}
\newcommand{\efg}{\end{figure}}

\begin{document}

\title{What is between Fermi-Dirac and Bose-Einstein Statistics?}

\author{Krzysztof Byczuk$^{a}$, Jozef Spa\l ek$^{b,a}$, Geoffrey Joyce$^{c}$, 
Sarben Sarkar$^c$}

\affiliation{(a) Institute of Theoretical Physics, Warsaw University, 
ul. Ho\.za 69, 00-681 Warszawa, Poland \\
(b) Institute of Physics, Jagellonian University, ul. Reymonta 4, 30-059
Krak\'ow, Poland \\
(c) Wheatstone Physics Laboratory, King's Collage, Strand, London WC2R 2LS,
UK}


\begin{abstract}
We overwiev the properties of a quantum gas of particles with the 
intermediate statistics defined by Haldane. Although this statistics
has no direct connection to the symmetry of the multiparticle wave
function, the statistical distribution function interpolates 
continuously between the Fermi-Dirac and the Bose-Einstein limits.
We present an explicit solution of the transcendental equation for
the didtribution function in a general case, as well as determine
the thermodynamic properties in both low- and high-temperature
limits.\\
\vspace{1cm}
Published in Acta Physica Polonica B {\bf 26}, 2167 (1995)
\end{abstract}
\maketitle
\section{Introduction }
Statistical properties of quantum many-body systems are determined by
calculating thermal averages with a proper
distribution function, which represents the mean number of particles
occupying each of the single-particle states.
For the three dimensional systems of noninteracting particles this
distribution function is given by either the Fermi-Dirac (FD) or the 
Bose-Einstein (BE) function,
depending on whether the spin of the particles is half-integer or 
integer, respectively \cite{huang}.

In the systems with either strong interactions or of low space dimensionality 
$(d<3)$ the situation changes drastically.
It turns out that the population of single particle quantum level
can be given by a function which is of neither FD, nor BE form.
The reason for that may be twofold.
First of all, in the space of two dimensions the symmetry of the many body 
wave function is not necessarily either even or odd \cite{leinnas}.
When two identical  particles are exchanged the wave function acquires
an arbitrary phase factor $e^{i \alpha}$, where $\alpha$ is so called
statistical
angle.
In other words, the proper symmetry group in this case is the braid group for
which  the irreducible representations are $\chi (\alpha) = e^{i \alpha}$.
The particles obeying the fractional statistics are called {\em anyons}.
On the other hand, one can imagine that the interactions between particles
are sufficiently strong, e.g. singular, that number of available 
effective single-particle occupancies of each of the
levels depends on the number of particles already present in that state.
This property can also lead to the departure from FD and BE 
statistics in arbitrary spatial dimensions \cite{haldane}.

There is no universal theory of statistical properties
of the systems described above.
However, there are some exactly solvable models of interacting particles for
which the momentum distribution functions are modified \cite{spalek}.
One of the best known cases is the one dimensional Calogero-Sutherland
model for
which the statistical properties of quasiparticles change from BE to FD 
limits as the interaction increases \cite{bernard}.

Having in mind those exactly solvable examples and the fact that, in general, 
the statistics of elementary excitations (quasiparticles) in correlated systems
can be arbitrary, we believe it is valuable to study the thermal properties
of particles with nonstandard distribution functions.
In particular, in this paper we address the question how thermodynamic
properties evolve when the statistics changes from BE to FD limits.

Our paper is organized as follows. In the first Section we introduce 
three interpolation schemes between BE and FD statistics.
Then, in Section 2 and 3 we discuss the high and the low temperature
properties of such systems.
In Section 5 we present general phase diagrams for those three systems of 
particles.

\section{Routes of interpolating
between Fermi-Dirac and Bose-Einstein distribution functions}

For noninteracting bosons the momentum distribution function  is
\be
\nee = \frac{1}{e^{\beta(\e_i-\mu)}-1},
\label{2.1}
\ee
whereas for ideal fermions we have
\be
\nee = \frac{1}{e^{\beta (\e_i - \mu)}+1}. 
\label{2.2}
\ee
In these functions, and hereinafter, $\e_i$ denotes the energy of a single
particle state $i$, $\beta = 1/k_BT$, where $T$ is the temperature, and $\mu$
is the chemical potential.
Probably, the simplest modification of  FD and BE functions can be
carried out phenomenologically in the following manner (route 1)
\be
\nega = \frac{1}{e^{\beta (\e_i - \mu)}+\gamma}, 
\label{2.3}
\ee
where $\gamma$ is an arbitrary real number in the range $(-1,1)$. 
The limit $-1$ corresponds to BE statistics and $\gamma=1$ to FD one.
For other allowed values of $\gamma$ we have an intermediate statistics.
In particular, the limit $\gamma=0$ corresponds to the classical,
i.e. Boltzman, distribution function.
Particles with arbitrary $\gamma$ statistics are called {\em $\gamma$-ons}.
We show later that this ad hoc proposal represents a high-temperature
approximation to a general distribution function.

As far as we know, it is impossible  to derive the distribution function
(\ref{2.3}) exactly \cite{ra,my}.
However, starting from the modified Hilbert-Fock space with creation and 
annihilation operators obeying so called $q$-mutator algebra,
i.e. $a_ia_j^+ - e^{i \alpha} a_j ^+a_i = 0$ for $i\neq j$, there is a way
to recover the function (\ref{2.3}) approximately where $\gamma = \cos \alpha$
\cite{my}.
It is also interesting to observe that the distribution function (\ref{2.3})
can be obtained if we suppose that the number of ways of putting $N_i$ 
particles into $G_i$ states is given by
\be
w_i = \frac{1}{(N_i)!}
\left( \frac{(G_i)!}{(G_i-\gamma N_i)!}
\right)^{1/\gamma}.
\label{2.2e}
\ee
Then, using the standard combinatoric approach to quantum statistics
\cite{huang}
we recover eq.(\ref{2.3}).

The thermodynamic potential $\Xi_{\gamma}(z,V,T)$ for $\gamma$-ons is 
given by
\be
\Xi_{\gamma}(z,V,T) = \prod_i (1+ \gamma z e^{\beta \e_i })^{1/\gamma},
\label{2.4}
\ee
where $z=e^{\beta \mu}$ is the fugacity.
In eqs.(\ref{2.2e},\ref{2.4}) we supposed that $\gamma \neq 0$.
The case of $\gamma = 0$ must be treated separately.
The equation  of state for such gas of ideal $\gamma$-ons is
\be
\frac{pV}{k_BT} = \ln \Xi_{\gamma}(z,V,T) = \frac{1}{\gamma} 
\sum_i \ln (1+ \gamma z e^{\beta \e_i }),
\label{2.5}
\ee
where $p$ is the pressure and $V$ is the volume of the system.
In order to eliminate $z$ from this formula we need another equation
which determines the total number of particles $N$
\be
N = z \frac{\partial }{\partial z} \ln \Xi_{\gamma}(z,V,T) =
\sum_i \frac{1}{ \frac{1}{z} e^{\beta \e_i} + \gamma} = \sum_i \nega.
\label{2.6}
\ee
In deriving the last equation  we could also convince ourselves
that the thermodynamic potential (\ref{2.4}) is really consistent
with the distribution function (\ref{2.3}).
Eqs. (\ref{2.5}) and (\ref{2.6}) describe the whole thermodynamics of 
$\gamma$-ons.
We are going to explore them in the next sections.

On the other hand, one can interpolate between FD and BE limits using
the state counting arguments. This provides us with the second 
possibility (route 2).
Namely, let the number of single particle state $D_i$ in the $i$-th 
quantum level depends on the number of particles in this group according to
the formula \cite{haldane,wu}
\be
\dd = G_i - g (N_i -1).
\label{2.7}
\ee
Also, suppose that the total number of states of N-body system for a given 
configuration $\{\n\} $ is \cite{haldane}
\be
W(\{ \n \}) = \prod_i 
\frac{(G_i - g(N_i -1) -1 + N_i)!}{(N_i)! (G_i - g(N_i-1)-1)!},
\label{2.8}
\ee
then using the Boltzman equation for the entropy
$S = k_B\ln \Gamma (\e)$, where $\Gamma (\e) = \sum_{\{\n\}} W(\{\n\})$ is
the total number of states,
we can derive the momentum distribution function $\negg$ \cite{wu}.
In eqs. (\ref{2.7}) and (\ref{2.8}) $G_i$ is the bare number of single 
particle states (when no particles are present).
$g$ is so called statistical interaction, or statistical parameter,
which in fact parameterize the distribution function.
We note that for $g=0$, the number of states (\ref{2.8}) 
is the same as for the ideal
bosons.
The case $g=1$ corresponds to free fermions.
However, for $0 \leq g \leq 1$, we can again obtain an arbitrary 
distribution function.
Physical interpretation of $g$ is follows.
Ideal bosons do not obey exclusion principle, so the number of states in the
$i$-th energy group would not change if we put another particle into this 
system and, therefore, $\dd = G_i$.
Fermions obey Pauli exclusion principle, which states that at most
only one fermion can
be present in a given quantum state.
Then, if we put another fermion into the $i$-th group the number of
accessible state  must decrease by one.
Hence, $\dd = G_i - N_i +1$ and $g=1$.
For arbitrary $g$ between zero and one we have a partial exclusion principle
and equation (\ref{2.8}) directly leads to fractional exclusion statistics
in arbitrary spatial dimensions.
Explicitly, the distribution function for any $g$ is given by the following
transcendental equation \cite{wu}
\be
\negg e^{\beta(\e_i-\mu)} = (1 + (1-g) \negg )^{1-g} (1- g \negg )^g.
\label{2.9}
\ee
Of course,  taking $g=0$ or $g=1$ in this equation we recover distribution 
functions (\ref{2.1}) and (\ref{2.2}).
Particles obeying  statistics with arbitrary $g$ are called {\em $g$-ons}.
From the last equation we find  that the distribution function can be
written in the following form
\be
\negg = \frac{1}{w_g(\e_i) + g},
\label{2.10}
\ee
where $w_g(\e_i)$ must satisfy the equation 
\be
e^{\beta (\e_i - \mu)} = (w_g(\e_i) + 1)^{1-g} (w_g(\e_i))^g.
\label{2.11}
\ee

As in the standard statistical mechanics \cite{huang}
we can find the thermodynamic potential
for  $g$-ons $\Xi_g(z,V,T)$ and, hence, the equation of state is
\be
\frac{pV}{k_BT} = - \sum_i \ln \left( 
\frac{1+(1-g)\negg}{1-g\negg}
\right).
\label{2.12}
\ee
The equation for the total  number of $g$-ons is
\be
N= \sum_i \frac{1}{w_g(\e_i) +g}.
\label{2.13}
\ee
The function $\negg$ and $w_g(\e_i)$ are given by eqs. (\ref{2.9}) and 
(\ref{2.11})
respectively.
Again, these functions (\ref{2.12}) and (\ref{2.13}) 
with (\ref{2.9}) and (\ref{2.11})
determine the whole thermodynamics of $g$-ons.

Finally, there is another  way of changing FD and BE statistics (route 3).
Namely, suppose that the number of particles in each single particle quantum
state can be equal to $p$, where $p$ is a natural number, i.e.
\be
N_i = 0,1,2,...p.
\ee
The case $p=1$ corresponds to free fermions and $p=\infty$ gives BE statistics.
The thermodynamic potential for a given $p$ is \cite{gentile}
\be
\Xi_p(z,V,T) = \prod_i \sum_{n_i = 0}^p 
e^{-\beta(\e_i-\mu)n_i} = \prod_i 
\frac{1-e^{(p+1)\beta(\e_i-\mu)}}{1-e^{-\beta(\e_i-\mu)}}.
\ee
To derive this we used the method of grand canonical ensemble.
With the aid of the property $\nep =  z \partial /\partial z \ln \Xi_p$
we can obtain the distribution function
\be
\nep = \frac{1}{e^{\beta (\e_i-\mu)}-1} -
\frac{p+1}{e^{(p+1)\beta(\e_i-\mu)}-1}.
\label{2.16}
\ee
The equation of state for a gas of so called {\em $p$-ons} is
\be
\frac{ pV }{ k_BT } = \sum_i \ln
\frac{ 1-z^{p+1} e^{ - \beta \e_i (p+1) } }{ 1 - ze^{- \beta\ e_i} },
\ee
and equation for their number is
\be
N = \sum_i \left(
 \frac{1}{e^{\beta (\e_i-\mu)}-1} -
\frac{p+1}{e^{(p+1)\beta(\e_i-\mu)}-1}
\right).
\ee

Let us summarize this Section.
We introduced three schemes of modified quantum statistics and, as a result,
we obtained three families of new nonstandard statistical distribution
functions which are parameterized by some numbers $(\gamma, g,p)$.
In these cases the statistics interpolate between FD and BE limits as the
statistical parameters are changed.
In the next sections we are going to discuss thermal properties of such
systems as a function of these parameters.

\section{High temperature properties}

Let us consider first the ideal $\gamma$-ons with parabolic dispersion 
relation $\e_k = \frac{\hbar^2 k^2}{2m}$ moving in the $d$-dimensional
space.
Then, we can expand the functions (\ref{2.5}) and (\ref{2.6}) in 
powers of $\gamma z$, and performing the simple Gaussian integrals we find 
that
\be
\frac{p}{k_BT} = \frac{1}{\lambda^d} \frac{1}{\gamma}
f_{\frac{d}{2}+1} (\gamma,z),
\label{3.1}
\ee
\be
n=\frac{N}{V} = \frac{1}{\lambda^d} \frac{1}{\gamma}
f_{\frac{d}{2}} (\gamma,z),
\label{3.2}
\ee
where the function $f_k (\gamma,z)$ is defined as 
\be
f_k(\gamma,z) = \sum_{i=1}^{\infty} (-1)^{i+1} \frac{(\gamma z)^i}{i^k}.
\ee
$\lambda = \sqrt{ 2\pi \hbar^2/mk_BT}$ is the thermal wavelength, and
$n$ is the particle density.
Of course, one can simply check that these equations reduce to those for
ideal fermions or bosons as $\gamma=1$ or $\gamma=-1$ respectively.

Now, considering the high temperature limit, i.e. a limit
for which $n\lambda^d \gamma \ll 1$, we can solve eq. (\ref{2.3}) 
finding  the fugacity 
\be
z \simeq n \lambda^d + \frac{\gamma (n \lambda^d)^2}{2^{d/2}}.
\ee
Then, substituting this into eq.(\ref{3.1}) we obtain the high-temperature 
(low-density) limit  of the equation of state
\be
\frac{p}{k_BT} = n (1 + A_2^{\gamma} n + \cdot \cdot \cdot),
\ee
where $A_2^{\gamma}$ is a second virial coefficient
\be
A_2^{\gamma} \equiv \gamma \frac{\lambda^d}{2^{d/2+1}}.
\ee
The character of the first quantum correction to the classical gas
depends on the sign of $\gamma$.
For $\gamma > 0$,
 $A_2^{\gamma}$ is positive and, effectively, particles repel
each  other.
In this case the pressure is higher than for the classical gas.
For $\gamma <0$ the second virial coefficient is negative
which means that the pressure is lowered.
In this range of statistical parameter $\gamma$ particles tend to collapse
because they effectively attract each other.
Roughly speaking, for $\gamma > 0$ the particles resemble fermions
and for $\gamma <0$ they look like bosons.
At the boundary ($\gamma =0$) between these two limits
there is no quantum correction to the classical behavior.

We can also calculate the internal energy $U=(d/2)pV$ and, hence, the specific 
heat of $\gamma$-ons at high temperatures.
The result is
\be
C_V = \frac{k_Bd}{2} n ( 1 + A_2^{\gamma} n (1-d/2) + \cdot \cdot \cdot).
\ee
In the one dimensional case the specific heat is decreasing function of 
temperature for positive $\gamma$.
For $d=3$ the behavior is opposite whereas, for $d=2$ the specific heat
 does not depend on temperature.
Also, we see that as $T\rightarrow \infty $ the specific heat goes to its
classical value $(d/2)R$, where $R$ is a gas constant ($R=nk_B$), and this
result is independent of $\gamma$.

Next, we examine the high temperature properties of $g$-ons.
The momentum distribution function for them is not given explicitly but
via the solution of eq. (\ref{2.9}).
However, we can find exactly this solution in term of
Taylor power expansion \cite{sarkar}.
To do this we solve eq. (\ref{2.9}) by 
utilizing the Legendre inversion theorem. 
The distribution function is in the form of the expansion
\be
\negg = \sum_{m=1}^{\infty} 
\frac{(mg+g-m)_m}{m!}
\frac{(-1)^m}{\xi_i^{m+1}},
\label{3.8}
\ee
where $\xi_i = e^{\beta (\e_i-\mu)}$ and $(a)_m = a(a+1)\cdot \cdot \cdot
(a+m-1) = \Gamma(a+m)/\Gamma(a)$.
For details see Appendix A.
Then it is straightforward to find the equation of state for ideal $g$-ons
\be
\frac{p}{k_BT} = \frac{1}{\lambda^d} f_{d/2 + 1}(g,z),
\label{3.9}
\ee
and the equation for the total number of particles
\be
n= \frac{N}{V} = \frac{1}{\lambda^d} f_{d/2}(g,z).
\label{3.10}
\ee
Again, $\lambda$ is the thermal wavelength and function $f_k(g,z)$ is defined
as follows
\be
f_k(g,z) = \sum_{m=0}^{\infty} (-1)^m \frac{(gm+g-m)_m}{m!}
\frac{z^{m+1}}{(m+1)^k}.
\ee
Eliminating $z$ from eq. (\ref{3.9}) with the aid of eq. (\ref{3.10}) 
we find the 
equation of state in the high temperature limit in the form
\be
\frac{p}{k_BT} = n (1 + A_2^g n + \cdot \cdot \cdot ),
\ee
where the second virial coefficient for $g$-ons is 
\be
A_2^g = (2g-1) \frac{\lambda^d}{2^{d/2+1}}.
\ee
For $g>1/2$ this coefficient is positive, so $g$-ons are like fermions.
However, for $g<1/2$ they resemble bosons in that sense that they tend to
condense.
We also see that identifying statistical  parameters $\gamma$ and $g$,
i.e.
\be
\gamma = 2 g -1,
\label{3.14}
\ee
these two sorts of particles obeying fractional statistics are 
identical at high temperatures.
Also, the whole results for the internal energy and the specific heat
for $\gamma$-ons could be repeated for $g$-ons if we use the relation 
(\ref{3.14}).

Finally, we examine the high temperature properties of $p$-ons.
Again, the distribution function (\ref{2.16}) can be expanded in power series
and Gaussian integrals can be performed easily for the ideal-gas case.
We find the equation of state and equation for particle 
number in the  forms
\be
\frac{p}{k_BT} = \frac{1}{\lambda^d} g_{d/2 +1} g(p,z),
\label{3.15}
\ee
\be
n= \frac{N}{V} = \frac{1}{\lambda^d} g_{d/2}(p,z).
\ee
In the present case the function $g_k(p,z)$ is following
\be
g_k(p,z) = \sum_{m=1}^{\infty}(-1)^{mp} \frac{z^{m(p+1)}}{m^k}
+\sum_{m=1}^{\infty} \frac{z^m}{m^k}.
\ee
If we take $p>1$ then the fugacity $z$ is 
\be
z \simeq n \lambda^d - \frac{(n\lambda^d)^2}{2^{d/2}},
\label{3.18}
\ee
where we dropped all higher order terms.
For $p=1$ the sign in the second term on the right hand side of eq. 
(\ref{3.18})
would be  positive.
Substituting (\ref{3.18}) into (\ref{3.15}) and taking only terms of the 
order $n^2$ we have
\be
\frac{p}{k_BT} = n( 1+ A_2^p n + \cdot \cdot \cdot),
\ee
where
\be
A_2^p = - \frac{\lambda^d}{2^{d/2+1}}.
\ee
We see that the second virial coefficient for $p$-ons does not 
depend on $p$ and is always negative for $p>1$.
Therefore, we conclude that at high temperatures these particles resemble
ordinary bosons.

To summarize this Section, we notice that for the first two statistical
distribution functions we obtained the same behavior in the 
high temperature limit, whereas in the third case the result was different.
In other words, $g$-ons and $\gamma$-ons behave similarly whereas $p$-ons 
always have the boson type corrections to classical limit.
In the next section we check the low temperatures properties of those 
systems.

\section{Low temperature properties}

Low temperature properties of $\gamma$-ons must be discussed separately
for  positive and  negative $\gamma$.
Let us consider first the $\gamma > 0$ case.

At $T=0$ the momentum distribution function reduces to the step function
\be
\negg = \frac{1}{\gamma} \theta (\mu - \e_i).
\label{4.1}
\ee
All states below the Fermi level are occupied with the mean population 
$1/\gamma$.
As $\gamma \rightarrow 0$ this number diverges.
For finite $\gamma$ we see that the averaged number of particles at each
quantum level could be  fractional and greater than one.
The states above the Fermi level are empty.

For free particles we can find the Fermi energy $\e_F^{\gamma} = \mu(T=0)$.
Namely, using  eq. (\ref{4.1}) we calculate that for a given density of 
particles
\be
\e_F^{\gamma} = \left( \frac{nd |\gamma|}{2 \Omega}\right)^{\frac{2}{d}},
\ee
where $\Omega = \frac{1}{(2 \pi)^{d/2} \Gamma(d/2)} (\frac{m}{\hbar^2})^{d/2}$.
The length of the Fermi vector 
$|{\bf k}_F|= \sqrt{ \frac{ 2 m \e_F^{\gamma} }{ \hbar^2 } }$ 
depends on the value  of $\gamma$ and so does 
the volume enclosed by the Fermi surface.
In particular we see that as $\gamma \rightarrow  0$ the Fermi vector 
disappears and all particles are on the lowest quantum level.

The existence of the Fermi surface implies that the low temperature 
properties of a system are affected only by the low energy 
particle-hole excitations close to it.
In other words, only the particles from the vicinity of the Fermi surface give
 contributions to the thermodynamics in this limit.
Therefore, we perform the low temperature (Sommerfeld type) expansion
for any thermal quantity. 
However, in order to do this we have to make some extra modification in 
the distribution function (\ref{2.3}).
Let us rewrite it as follows
\be
\frac{1}{e^{\beta (\e - \mu)}+\gamma} = 
\frac{1}{\gamma} \frac{1}{e^{\beta (\e - \mu^{\ast})}+1},
\ee
where $\mu^{\ast} \equiv \mu + k_BT \ln \gamma$.
Now, we can expand the thermodynamic functions around the point $\mu ^{\ast}$.
The general scheme of calculation is presented in Appendix B.
Here we quote only the final results.

The lowest corrections to the chemical potential are 
\be
\mu = \e_F^{\gamma} - k_BT \ln \gamma - \frac{\pi^2}{6} 
\frac{d-2}{2 \e_F^{\gamma}} \frac{1}{|\gamma|}
 (k_BT)^2.
\ee
For ordinary fermions $(\gamma=1)$ we had the first correction  to the 
chemical potential proportional to $T^2$.
Now, the linear in temperature correction has appeared.
This is a direct consequence of the modified form of the distribution 
function (\ref{2.3}).
However, this term disappears in the formula for the internal energy.
Namely, as was  shown in Appendix B the internal energy is
\be
\frac{U}{V} = \bar{\e}_{\gamma}
+ \frac{\Omega}{|\gamma|} \left( \frac{n d |\gamma|}{2 \Omega}
\right)^{1 - \frac{2}{d}}
 (k_BT)^2.
\ee
The general form is similar to that for fermions.
The only difference is the factor $1/\gamma$ which enhances the density 
of states at the Fermi level.
Also, the specific heat 
\be
C_V = K_{\gamma} T
\ee
is linear function of $T$.
The coefficient $K_{\gamma} \sim (1/ \gamma)^{2/d}$ 
diverges as $\gamma \rightarrow 0$.

So, we proved that the ground state and the low temperatures 
properties of $\gamma$-ons for $\gamma >0$ resemble properties
of ordinary fermions.
However, it is not the case for $\gamma < 0$.

For negative statistical parameter $\gamma$ the distribution function
(\ref{2.3}) is singular when 
$\e_i + k_BT \ln \gamma - \mu =0$.
This may be a signal of  the Bose-Einstien condensation (BEC)
at least in dimensions greater than two.
Let us consider this case.

The equation for total number of particles is 
\be
n=n_0 + \frac{1}{\lambda ^d |\gamma|} g_{1/d + 1} (|\gamma|,z),
\ee
where $n_0$ is the number of particles at the first quantum level (with
${\bf k} =0$) divided by $V$. 
The function $g_{k}$ is given by
\be
g_k (|\gamma|,z) = \sum_{m=1}^{\infty} 
\frac{(|\gamma|z)^m}{m^k}.
\ee
BEC occurs when $|\gamma|z=1$ and then the state ${\bf k}=0$ starts to be 
microscopically occupied.
This gives that in three dimensional space we have the following 
condition \cite{huang}
\be
\lambda^3 n | \gamma| = 2.612.
\ee
Hence, we find the critical temperature 
\be
k_BT_c^{\gamma} = \frac{2 \pi \hbar ^2}{m} 
\left(
\frac{n |\gamma|}{2.612}
\right)^{2/3}.
\label{4.11}
\ee
This temperature is smaller for particles with fractional bosonic statistics
than for ordinary bosons.
Below $T_c$ the contribution to the internal energy comes from particles
which are outside the condensate and we can easily find that 
\be
\frac{U}{V} = 
\frac{\Gamma (d/2 + 1) \xi (d/2 +1)}{\Gamma(d/2) \xi(d/2)}
n (k_BT) \left( \frac{T}{T_c^{\gamma}}
\right)^{d/2}.
\label{4.12}
\ee
Hence, the specific heat is 
\be
C_V = \frac{\Gamma (d/2 + 1) \xi (d/2 +1)}{\Gamma(d/2) \xi(d/2)}
\left(\frac{d+2}{2}
\right) k_B \left( \frac{T}{T_c^{\gamma}}
\right)^{d/2},
\ee
and the equation of state is 
\be
p= \frac{d}{2} \frac{\Gamma (d/2 + 1) \xi (d/2 +1)}{\Gamma(d/2) \xi(d/2)}
n (k_BT) \left( \frac{T}{T_c^{\gamma}}
\right)^{d/2}.
\label{4.14}
\ee
The dependence on $\gamma$ is hidden  in $T_c^{\gamma}$ only.
On eq. (\ref{4.11}) we see that as $|\gamma| \rightarrow 0$ then $T_c^{\gamma}
\rightarrow 0$ too, and hence, the thermal functions (\ref{4.12}) - 
(\ref{4.14})  diverges.
This expresses the fact that the limit $\gamma = 0$ is a nonanalytic 
boundary between particles with $\gamma >0$ and with $\gamma < 0$.
In other words, this particular value of  $\gamma$ establishes a true phase
boundary between two phases of noninteracting $\gamma$-ons.

In the next part we are going to discuss the low temperature properties of 
$g$-ons. 
At $T=0$ the momentum distribution function reduces to 
\be
\negg = \frac{1}{g} \Theta (\mu - \e_i),
\ee
for all $0<g\leq 1$. 
This means that for all positive $g$ we have a true Fermi surface in the 
reciprocal space which separates occupied and unoccupied states.
Note also, that the case $g=0$ must be treated separately because then
$\negg $ is infinite.
Again, as in the case of $\gamma$-ons with $\gamma >0$, the main 
contribution  to the low energy properties is given by
the particle-hole  excitations around the Fermi surface.
Performing the Sommerfeld-type expansion we find that (see Appendix B and also
\cite{wilczek})
\be
\mu = \e_F^g - \frac{d-2}{2\e_F^g} g C_1(g)  (k_BT)^2.
\ee
In the present case there is no linear  temperature corrections
to the chemical potential.
We also calculate the internal energy
\be
\frac{U}{V} = \bar{\e}_g
+ \Omega (\e_F^g)^{1-2/d}  C_1(g) (k_BT)^2,
\ee
where now the Fermi energy is
\be
\e_F^g = \left( \frac{ndg}{2 \Omega}\right)^{\frac{2}{d}}.
\ee
Hence, we obtain the specific heat 
\be
C_V = K_g T,
\ee
where 
$K_g \sim (1/g)^{2/d-1}C_1(g)$.
The low temperature properties of $g$-ons resemble those of ideal fermions with
modified density of states by the factor $1/g$. 
We also see that the case $g=0$ can not be reached analytically from the 
case $g>0$.
This is due to the possibility of BEC of ideal bosons ($g=0$).

As an additional problem it was interesting to consider BEC for ideal $g$-ons
with $g<0$.
As we proved rigorously in the Appendix C, there is no BEC in this case.
So, only true bosons can develop a macroscopic occupation of the 
lowest (${\bf k}=0$) quantum level at finite temperatures.

Finally, we consider $p$-ons at low temperatures.
At $T=0$ the distribution function (\ref{2.16}) is
\be
\nep = p \Theta (\mu - \e_i).
\ee
We see that, although both functions in eq. (\ref{2.16}) diverge
at $\e_i = \mu$ 
the final result is well behaved; those two singularities cancel 
each other.
The last result means that $p$-ons possess a sharp Fermi surface 
with a wavevector 
\be
|{\bf k}_F^p | = \sqrt{\frac{2m}{\hbar^2}}\left( \frac{nd}{2\Omega p}\right)^{1/d}
\ee
This property allows us to perform again the Sommerfeld (low temperature) 
expansion for thermodynamic quantities.
This was done in Appendix B. 
We note that for any $ 1 \leq p < \infty$ the specific heat is 
linear with temperature, i.e. 
\be
C_V = K_p T,
\ee
where now $K_p \sim \frac{p}{p+1} p^{2/d-1}$.
So, we conclude that at very low temperatures $p$-ons resemble free 
fermions with modified density of states.

\section{Conclusions}

In this paper we studied the thermodynamic properties of particles 
obeying three different statistics. Each of them interpolates the 
distribution function between the BE and the FD limits.
We compared both the high and the low temperature properties of 
$\gamma$-ons, $g$-ons and $p$-ons.
We showed that $\gamma$-ons behave like fermions when $\gamma >0$ and like 
bosons when $\gamma <0$ at all temperatures.
However, their properties are not the same
because their thermal functions are parameterized by $\gamma$
which changes the number of states.
We also showed that in between of the boselike and the fermilike regimes,
there is a true boundary condition $(\gamma=0)$, where the statistics
is classical.
However, when we approached this limit from the left or the right sides
the thermal functions diverge as $1/\gamma$.
In fig. 1 we draw the schematic phase diagram for $\gamma$-ons for 
different values of $\gamma$.

The thermodynamic properties of $g$-ons turned out to be quite 
different.
At high temperatures we noted two classes of behavior: fermilike for $g>1/2$
and boselike for $g<1/2$.
For $g=1/2$ the statistics was classical.
However, at low temperatures $g$-ons resemble ordinary fermions with 
modified density of states for  all $0<g\leq 1$.
This means that there must be a crossover temperature when 
the typical behavior changes its character.
The schematic phase diagram is provided in fig. 2.

Finally, we considered $p$-ons and we showed that for $p>1$ they look like 
bosons at very high temperature.
On the other hand, they resemble fermions at low temperatures if $p < \infty$.
So, again there must exist a crossover temperature from the bosonic to 
the fermionic behavior.
The phase diagram for $p$-ons is shown in fig. 3.

In conclusions, we showed that different routes of
interpolating between FD and BE
limits corresponds to different ways of approaching the 
statistical mechanics of particles with
a nonstandard statistics. Our methods 2 and 3 are based on the combinatorial
approach, which is not directly related to the symmetry properties of
the many-body wave function with respect to the particle transposition.
In particular, the knowledge of the second virial coefficient is not enough
to determine the full thermodynamics. For example, although we know the 
virial coefficient for anyons, it is impossible distinguish which statistics
describes them at low temperatures, $\gamma$-ons or $g$-ons.
We think that none of them is the proper one.
However, anyons in the strong magnetic field seem to be described by
$g$-on statistical mechanics \cite{can}.

\section{Appendix A}

In this Appendix we show how to solve eq. (\ref{2.9}) using the 
Lagrange inversion theorem.
Suppose, that we have a transcendental equation 
\be
y = a + x \phi (y),
\label{aa1}
\ee
and we want to find $y$ as an unknown variable.
The variables $x$ and $a$ are supposed to be given.
The solution can be represented by infinite power 
series
\be
y = a + \sum_{n=1}^{\infty} \frac{x^n}{n!} 
\frac{d^{n-1}}{dy^{n-1}} \left[ \phi^n(a)
\right].
\label{aa2}
\ee
Noting, that eq. (\ref{2.9}) has the same formal structure as (\ref{aa1}) with
$a=0$ we use eq. (\ref{aa2}) to obtain eq. (\ref{3.8}).

\section{Appendix B}

In this Appendix we present the low temperature expansion for any thermodynamic
function calculated with arbitrary statistics. The only limitation
is that this thermal function must be well behaved in the vicinity of 
chemical potential.
Let us consider the following integral
\be
I = \int_0^{\infty} H(\e) n(\e - \mu) d \e
\ee
Introducing the new variable $y = \beta \e - \beta \mu$ and manipulate with 
the integrals we obtain
\begin{widetext}
\be
I = \frac{1}{\beta} \int_0^{\infty} dy H(\frac{y}{\beta} + \mu) n(y)
+\frac{1}{\beta} \int^0_{-\beta \mu} dy H(\frac{y}{\beta} + \mu) (n(y) - a)
+\frac{1}{\beta} \int^0_{-\beta \mu} dy H(\frac{y}{\beta} + \mu) a ,
\ee
\end{widetext}
where the constant $a$ must be chosen properly to the statistics. For example,
for $\gamma$-ons we should take $a=1$, $g$-on statistics needs $a=1/g$ whereas
for $p$-ons $a=p$. At low temperatures we can approximate the second integral
extending the lower limit of integration to $-\infty$. Then, changing the 
variable in this integral $y \rightarrow -y$ and expanding both functions
$H(\mu \pm k_BT y)$ around the point $\mu$ we obtain
\be
I = a \int_0^{\mu} H(\e) d\e + \sum_{i=0}^{\infty}
C_i(a) \frac{1}{i!} \frac{d^iH(\mu)}{d\e^i} (k_BT)^{i+1},
\label{a}
\ee
where the coefficients $C_i(a)$ are defined as follows
\be
C_i(a) = \int_0^{\infty} dy y^i (n(y) + (-1)^i (n(-y) - a)).
\ee 

If $a=1$ and $n(y)$ is $\gamma$-ons
 momentum distribution function we obtain that
the  even coefficients
 are zero $C_{2n}(1)=0$ and the odd coefficients are given by
\be
C_{2n-1} (1) = (-1)^{n-1} (2 \pi)^{2n} \frac{B_{2n}}{2n} (1 - 2^{1-2n}),
\ee
where $B_{2n}$ are the Brilouin numbers. So, $\gamma$-ons have almost 
the same  expansion as ideal fermions.

Similar results we obtain for $p$-ons. Using the distribution function
(\ref{2.16}) we find that
\be
C_{2n-1} (p) = (-1)^{n-1} (2 \pi)^{2n} \frac{B_{2n}}{2n} ( 1 - (p+1)^{1-2n})
\ee
and $C_{2n}=0$. Note that for $p=\infty$ we obtain the coefficients for the 
Sommerfeld expansion of ideal bosons provided that they do not condense.

For $g$-ons it is impossible to find the analytic formula for $C_k(g)$.
We only know that $C_0(g)=0$ \cite{wilczek}
 and the rest ones are finite.

The formalism  developed above is useful in calculating the chemical
potential and the internal energy. Namely, the equation for $\mu$ is
\be
n=\frac{N}{V} = \Omega \int_0^{\infty} d \e \e^{d/2 -1} n(\e),
\label{a1}
\ee
and for the energy
\be
\frac{E}{V} = \Omega \int_0^{\infty} d\e \e^{d/2} n(\e),
\label{a2}
\ee
where
\be 
\Omega = \frac{1}{(2\pi)^{d/2}\Gamma(d/2)} \left(
\frac{m}{\hbar^2}
\right)^{d/2}.
\ee
Expanding eq. (\ref{a1}) to first nontrivial order as in eq. (\ref{a})
we have
\be
n \simeq \Omega ( a \int_0^{\mu} d\e \e^{d/2-1} + 
C_1(a) (d/2 -1 ) (\mu)^{d/2 -2} (k_BT)^2),
\ee
and then supposing that the correction to the chemical potential is small
$\mu = \e_F^a + \delta \mu$ we can find that
\be
\mu = \e_F^a \left( 1 - (d/2 -1) 
\frac{C_1(a)}{a} \left( \frac{k_BT}{\e_F^a}
\right)^2 \right).
\ee

Similarly, we find the low temperature internal energy
\be
\frac{E}{V} = \bar{\e} + \Omega (\e_F^a)^{d/2 -1} C_1(a) (k_BT)^2,
\ee
where $\bar{\e}_a= \frac{2 \Omega a}{d+2} (\e_F^a)^{d/2 +1}$.
The last equation allows us to calculate the specific heat which is
\be
C_V = K_a T,
\ee
where $K_a= 2 \Omega (\frac{nd}{2a\Omega})^{1-2/d} C_1(a) k_B^2$

\section{Appendix C}

In this Appendix we extend our discussion of $g$-ons to the case with 
negative $g$ (i.e. we suppose that $g \leq 0$).
In principle, the value of $g$ can be arbitrary and  it seems very interesting
to investigate how the particles look like with "{\em attractive}"
statistical interaction.
The most exciting question deals with the BEC in such system.
However, as we show below, the condensation does not occur in such gas.

The equation for the momentum distribution function is following
(cf. (\ref{2.9}))
\be
\negg \xi = (1 + (1 + \eta)\negg)^{1+\eta} ( 1 + \eta \negg)^{-\eta}, 
\label{c1}
\ee
where we define $g = - \eta$, with $\eta >0$, and $\xi = e^{\beta (\e_i-\mu)}$.
Using the method developed in ref. \cite{sarkar} (see also Appendix A)
we find the solution of eq. (\ref{c1}) in the form of the infinite power
series, i.e.
\be
\bar{n}_g(\xi) = \sum_{m=0}^{\infty}
\frac{(m \eta + \eta + m)_m}{m!}
\frac{1}{\xi^{m+1}}.
\ee
This series is absolutely converged for $\xi > \frac{(\eta+2)^{\eta+2}}{
(\eta+1)^{\eta+1}}$.
Next, we find the equation for the total particle number 
\be
\frac{N}{V} = \frac{1}{V} \bar{n}_0 + \frac{1}{\lambda^d}
g_{d/2}(\eta,z),
\label{c2}
\ee
where the function $g_k(\eta,z)$ is given by the following series
\be
g_k(\eta,z) = \sum_{m=0}^{\infty}
\frac{(\eta m + \eta + m)_m}{m!} 
\frac{z^{m+1}}{(m+1)^{k}}.
\ee
$\bar{n}_0$ is the population of the lowest energy level and we treat it 
separately in case of BEC.
Eq. (\ref{c2}) is absolutely converged for all $z$ in the range
\be
0 \leq z \leq z_R \equiv \frac{(\eta+1)^{\eta+1}}{(\eta+2)^{\eta+2}}.
\ee
On the other hand, momentum distribution function can be written in the 
form (\ref{2.10}) where $w(\xi)$ obeys the equation
\be
w^{-\eta} ( 1 + w)^{1 + \eta} = \frac{1}{z} e^{\beta \e}.
\ee
Condition for BEC is $w=\eta$. Substituting this into eq. (\ref{c1}) and
taking the maximal value of possible $z=z_R$ we obtain the allowed $\e$
\be
\beta \e = 2 (1+\eta)\ln (1+\eta) - (2+\eta)\ln (2+\eta) - \eta \ln \eta.
\ee
We see that in order to have BEC $\e$ must be negative but we have
 supposed that $\e \geq 0$, so we obtain contradiction and this proves 
the absence of BEC for $g$-ons.

\end{document}